\documentclass[aps,prl,twocolumn,floats,final,superscriptaddress,citeautoscript,longbibliography]{revtex4-2}
\usepackage{multirow}
\usepackage{hhline}
\usepackage{enumerate}
\usepackage{amssymb,bm}
\usepackage{graphicx}
\usepackage{amsmath}
\usepackage{braket}
\usepackage{mathtools}
\usepackage{bbm}
\usepackage[colorlinks=true,linkcolor=blue,citecolor=blue,urlcolor=blue]{hyperref}
\usepackage{xcolor,soul}
\usepackage{changes}
\usepackage{lineno,comment}

\begin{document}

\title{Contractive Unitary and Classical Shadow Tomography}


\author{Yadong Wu}
\affiliation{College of Physics, Sichuan University, Chengdu 610065, China}
\affiliation{Department of Physics and Hong Kong Institute of Quantum Science and
Technology, The University of Hong Kong, Pokfulam Road, Hong Kong SAR, China}

\author{Ce Wang}
\affiliation{School of Physics Science and Engineering, Tongji University, Shanghai, 200092, China}

\author{Juan Yao}
\affiliation{Shenzhen Institute for Quantum Science and Engineering, Southern University of Science and Technology, Shenzhen 518055, Guangdong, China}
\affiliation{International Quantum Academy, Shenzhen 518048, Guangdong, China}
\affiliation{Guangdong Provincial Key Laboratory of Quantum Science and Engineering, Southern University of Science and Technology, Shenzhen 518055, Guangdong, China}

\author{Hui Zhai}
\affiliation{Institute for Advanced Study, Tsinghua University, Beijing 100084, China}
\affiliation{Hefei National Laboratory, Hefei 230088, China}

\author{Yi-Zhuang You}
\affiliation{Department of Physics, University of California at San Diego, La Jolla, CA 92093, USA}

\author{Pengfei Zhang}
\email{pengfeizhang.physics@gmail.com}
\affiliation{Department of Physics \& State Key Laboratory of Surface Physics, Fudan University, Shanghai 200438, China}
\affiliation{Shanghai Qi Zhi Institute, AI Tower, Xuhui District, Shanghai 200232, China}

\date{\today}

\begin{abstract}
The rapid development of quantum technology demands efficient characterization of complex quantum many-body states. However, full quantum state tomography requires an exponential number of measurements in system size, preventing its practical use in large-scale quantum devices. A major recent breakthrough in this direction, called classical shadow tomography, significantly reduces the sample complexity, the number of samples needed to estimate properties of a state, by implementing random Clifford rotations before measurements. Despite many recent efforts, reducing the sample complexity below $\bm{2^k}$ for extracting any non-successive local operators with a size $\sim \bm{k}$ remains a challenge. In this work, we achieve a significantly smaller sample complexity of $\bm{\sim 1.8^k}$ using a protocol that hybridizes locally random and globally deterministic unitary operations. The key insight is the discovery of a deterministic global unitary, termed as \textit{contractive unitary}, which is more efficient in reducing the operator size to enhance tomography efficiency. The contractive unitary perfectly matches the advantages of the atom array quantum computation platform and is readily realized in the atom array quantum processor. More importantly, it highlights a new strategy in classical shadow tomography, demonstrating that a random-deterministic hybridized protocol can be more efficient than fully random measurements.
\end{abstract}

\maketitle

Recent remarkable breakthroughs in quantum science have enabled high-precision preparation, manipulation, and measurement of highly entangled quantum wave functions involving a large number of qubits \cite{Arute:2019aa,science.abg5029,PRXQuantum.2.017003,PhysRevLett.127.180501,Acharya:2023aa,Evered:2023aa,Ma:2023aa,Bluvstein:2024aa}. However, characterizing complex quantum states poses a significant challenge because the exponential growth of the Hilbert space dimension demands exponential resources for a complete tomography of quantum states \cite{Flammia_2012,odonnell2015,7956181}, making it impractical for modern quantum devices with more than hundreds of qubits. Nevertheless, it has been discovered that full tomography is not necessarily required if our primary interest lies in physical properties, such as the expectations of Pauli string observables. Classical shadow tomography can predict many properties using a much smaller number of measurement records \cite{aaronson2018,paini2019,Huang:2020aa}. This strategy has attracted considerable attention and has become an important development at the forefront of modern quantum science \cite{PRXQuantum.2.010307,PRXQuantum.2.030348,PhysRevLett.127.110504,doi:10.1126/science.abk3333,PhysRevResearch.4.013054,Elben:2023aa,zhou2023,PhysRevResearch.6.013029,PhysRevB.109.094209,PRXQuantum.5.010350,PRXQuantum.5.020304,PhysRevResearch.5.023027,Akhtar2023scalableflexible,akhtar2024,2024arXiv240217911H,2024arXiv240611788Z,PhysRevLett.133.020602,PhysRevLett.130.230403,Bu:2024aa}.

\begin{figure*}
    \begin{center}  
    \includegraphics[width=0.8\linewidth]{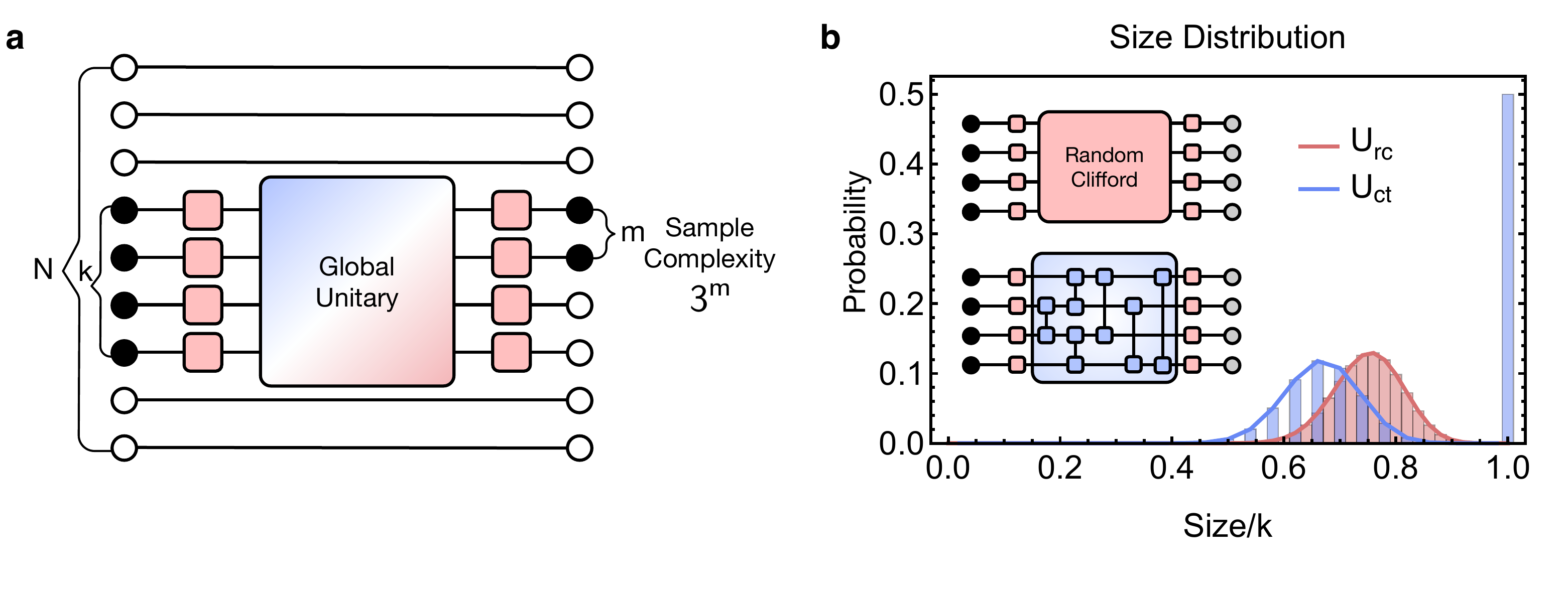}
    \caption{\textbf{Protocols of the classical shadow tomography and operator size distribution.}
    \textbf{a}, Schematics of the classical shadow tomography protocol for a Pauli string operator acting on a subsystem with $k$ qubits. The protocol comprises a global unitary sandwiched between two layers of random single-qubit rotations (red blocks) applied to the subsystem. If the unitary reduces the operator size of the Pauli string from size $k$ to size $m$, the sample complexity for predicting the Pauli operator is $3^m$. \textbf{b}, A comparison of the operator size distribution after applying a random Clifford circuit ($U_{\rm rc}$) or a contractive unitary ($U_{\rm ct}$, defined in the main text) to an input size-$k$ Pauli string operator. We choose an operator size of $k=50$ as an illustration. The size distribution under random Clifford evolution (red line) is broad, peaking at $3k/4$. In contrast, the size distribution under contractive evolution (blue line) exhibits a broad distribution peaking at $2k/3$ and a delta peak at $k$. The smaller peak value of the broad distribution for the contractive unitary is crucial for improving the efficiency of classical shadow tomography.  
    \label{Fig:illustration}}
    \end{center}
\end{figure*}

Fig. \ref{Fig:illustration}\textbf{a} schematically represents a broad class of classical shadow tomography protocols. We first consider the Pauli string operator $\hat{O}$ acting on a given subsystem $S$ consisting of $k$ qubits. The typical quantum circuit for classical shadow tomography contains a generic global unitary $\hat{U}_\text{g}$ sandwiched between two layers of random single-qubit rotations $\prod_i \hat{u}_{1,i}$ and $\prod_i \hat{u}_{2,i}$, independently sampled from the Clifford group. The random single-qubit rotations eliminate local basis dependence, ensuring that all operators can be measured on equal footing \cite{PhysRevResearch.5.023027,Bu:2024aa,Akhtar2023scalableflexible,akhtar2024,2024arXiv240217911H,2024arXiv240611788Z,PhysRevLett.130.230403,PhysRevLett.133.020602}. After applying the composite unitary operation $\hat{U}\equiv(\prod_i \hat{u}_{2,i})~\hat{U}_\text{g}~(\prod_j \hat{u}_{1,j})$, the subsystem is measured in the computational basis. Each measurement outcome $\ket{\mathbf{z}} \equiv\ket{z_1,\dots,z_k} $ with $z_j\in\{0,1\}$ provides a classical snapshot of the density matrix as $\hat{\sigma}_U(\mathbf{z})=\hat{U}^\dagger |\mathbf{z}\rangle \langle \mathbf{z}|\hat{U}$ \cite{Huang:2020aa}. 

A collection of enough classical snapshots allows us to predict the expectation value of all Pauli string operators $\hat{O}$ with size $\sim k$ by averaging the predictions under these classical snapshots as $\langle O\rangle=\Vert\hat{O}\Vert_{\mathcal{E}_{U}}^2~\mathbb{E}[\text{tr}(\hat{O} \sigma_U(\mathbf{z}))]$ \cite{Bu:2024aa}. Here, the key quantity $\Vert\hat{O}\Vert_{\mathcal{E}_{U}}^2$ is known as the shadow norm \cite{Huang:2020aa}, which depends on the choice of the unitary ensemble $\mathcal{E}_U$ and needs to be determined theoretically as a prior. Intuitively, if $\Vert\hat{O}\Vert_{\mathcal{E}_{U}}^2$ is larger, one needs to average over more classical snapshots to reduce the variance in predicting $\langle O\rangle$. In this sense, the shadow norm characterizes the sample complexity in classical shadow tomography. Furthermore, it can be shown that the variance of the predictions from classical snapshots is precisely given by the shadow norm. Therefore, the essential point is to find an optimal $\hat{U}$ with a smaller shadow norm to reduce the sample complexity and suppress the prediction variance.

Previous studies have also shown that the shadow norm is closely related to the operator size distribution of the evolved operator $\hat{O}_U\equiv \hat{U} \hat{O} \hat{U}^\dagger$ as follows \cite{qi2019,PhysRevLett.130.230403,Bu:2024aa}.
\begin{equation}\label{eqn:norm_to_size}
\Vert\hat{O}\Vert_{\mathcal{E}_{U}}^2=w(\hat{O}_U)_{\mathcal{E}_{U}}^{-1}, ~~ w(\hat{O}_U)_{\mathcal{E}_{U}}=\sum_m\frac{\pi(m)_{\mathcal{E}_{U}}}{3^{m}},
\end{equation}
where $w(\hat{O}_U)$ is known as the Pauli weight and $\pi(m)$ is the size distribution of $\hat{O}_U$ \cite{Roberts:2015aa,Roberts:2018aa}. For a Pauli string operator, the size $m$ is defined as the number of non-identity elements in the string. More generic operators can be expanded in the Pauli basis $\hat{O}_U=\sum_{\hat{P}}c_{\hat{P}}\hat{P}$ and thereby possess a size distribution $\pi(m)=\sum_{\text{Size}(\hat{P})=m} |c_{\hat{P}}|^2$. In the simplest case, if we choose $\hat{U}_\text{g}$ as the identity, the unitary does not change operator size, and a Pauli string operator with size $k$ possesses $\Vert\hat{O}\Vert^2= 3^k$. Another example of $\hat{U}_\text{g}$ is to randomly sample $\hat{U}_\text{rc}$ from the $k$-qubit Clifford group, such that the output operator is maximally scrambled. In this case, the size distribution follows a binomial distribution, as illustrated by the red line in Fig. \ref{Fig:illustration}\textbf{b}. Using Eq. \eqref{eqn:norm_to_size}, such a size distribution leads to a shadow norm $\Vert\hat{O}\Vert^2_{\text{rc}} \sim 2^k$. 

Eq. \eqref{eqn:norm_to_size} illustrates a general trend: if the evolved operator $\hat{O}_U$ has a size distribution peaked at a smaller operator size, the shadow norm will be smaller. A challenging question is whether there exist other choices of global unitaries that can outperform the maximally scrambled random unitary and result in a smaller shadow norm compared to $\sim 2^k$. The main result of this work is that we have discovered a deterministic unitary that can achieve better performance than the maximally scrambled random unitary in classical shadow tomography. We term this unitary as \textit{Contractive Unitary} because it can more efficiently contract operator size. We show that, by using the contractive unitary as $U_\text{g}$, we can achieve a shadow norm that scales as $1.8^k$, significantly reducing measurement resources when $k$ is large.

\textbf{Contractive Unitary.} We first motivate the construction of the contractive unitary by focusing on a subsystem $S$ with two qubits. There are nine size-2 Pauli operators in this subsystem. It can be shown that, under the evolution of a Clifford gate, at most four of them can be contracted to size-1 operators (see the Supplemental material \cite{supp}). We call a Clifford gate that saturates this bound the contractive unitary for the two-qubit case. An explicit construction of a two-qubit contractive unitary is:
\begin{equation}
\hat{U}_{12}=\exp\left(i\frac{\pi}{4}\hat{Z}_1\hat{Z}_2\right), \label{Uij}
\end{equation}
which maps four size-2 operators $\hat{X}_1\hat{Z}_2$, $\hat{Y}_1\hat{Z}_2$, $\hat{Z}_1\hat{X}_2$ and $\hat{Z}_1\hat{Y}_2$ to size-1 operators $-\hat{Y}_1\hat{I}_2$, $\hat{X}_1\hat{I}_2$, $-\hat{I}_1\hat{Y}_2$ and $\hat{I}_1\hat{X}_2$, respectively, while leaving the remaining five size-2 operators unchanged. Such a contractive unitary also approaches the optimal shadow norm of the two-qubit locally-scrambled unitary ensemble from the perspective of entanglement features \cite{akhtar2024}.

Two-qubit contractive unitaries are not unique. However, this particular choice has the following useful properties when calculating the sample complexity in the many-qubit case discussed below. A Pauli string operator $\hat{O}_1\hat{O}_2$ evolved by $\hat{U}_{12}$ falls into one of the following three situations, up to an unimportant phase factor:
\begin{enumerate}[{1)}]
\item If both $\hat{O}_1$ and $\hat{O}_2$ belong to $\{\hat{Z},\hat{I}\}$, the operator does not change. 

\item If both $\hat{O}_1$ and $\hat{O}_2$ belong to $\{\hat{X},\hat{Y}\}$, the operator does not change. 

\item If $\hat{O}_1$ belongs to $\{\hat{Z},\hat{I}\}$ and $\hat{O}_2$ belong to $\{\hat{X},\hat{Y}\}$, or vice versa, the evolution by $\hat{U}_{12}$ permutes $\hat{X}\leftrightarrow\hat{Y}$ and $\hat{Z}\leftrightarrow\hat{I}$.
\end{enumerate}

For a generic subsystem $S$ with $k$ qubits, we construct the contractive unitary as $\hat{U}_{\text{ct}}=\prod_{i<j}\hat{U}_{ij}$, where $\hat{U}_{ij}$ is given by Eq. \eqref{Uij}, acting on any pair of qubits $i$ and $j$ within the subsystem. Since $\hat{U}_{ij}$ gates with different $i$ and $j$ commute with each other, their ordering is not important. We first focus on size-$k$ Pauli string operators, which are tensor products of $\hat{X}$, $\hat{Y}$, or $\hat{Z}$ operators within the operator supports. Let $N_{XY}$ denote the total number of $\hat{X}$ and $\hat{Y}$ operators in the string operator, with the number of $\hat{Z}$ operators being $k - N_{XY}$. If the operator at site $i$ is $\hat{X}$ or $\hat{Y}$, it remains $\hat{X}$ or $\hat{Y}$ upon evolution by $\hat{U}_{\text{ct}}$, according to points 1)–3) above. However, if the operator at site $i$ is $\hat{Z}$, since $\hat{U}_{\text{ct}}$ connects site $i$ to all other sites ergodically, it flips $N_{XY}$ times between $\hat{Z}$ and $\hat{I}$ according to point 3). Thus, if $N_{XY}$ is odd, all $\hat{Z}$ operators are replaced by $\hat{I}$, reducing the operator size from $k$ to $N_{XY}$; if $N_{XY}$ is even, the $\hat{Z}$ operators remain unchanged. Therefore, we find 
\begin{equation}
m=\text{Size}\left(\hat{U}_{\text{ct}} \hat{O}\hat{U}_{\text{ct}}^\dagger\right)=
\begin{cases}
                   N_{XY} & \text{if}\ \   N_{XY} \in  \text{odd},  \\
                    k & \text{if}\ \   N_{XY} \in  \text{even}.
\end{cases}\label{eqn:size-ct}
\end{equation}
In Fig. \ref{Fig:illustration}\textbf{b}, we present the operator size distribution $\pi(m)$ using a bar chart for a subsystem with $k=50$. It contains two contributions: a broad peak near $m/k\approx 2/3$ due to odd $N_{XY}$ and a delta peak at $m/k=1$ due to even $N_{XY}$. The total weights of these two contributions are equal. In contrast, the size distribution of a maximally scrambled operator peaks around $3/4$, as also shown in Fig. \ref{Fig:illustration}\textbf{b}. It is interesting to note that this contractive unitary reduces the operator size more efficiently than the random unitary for half of the Pauli string operators, though at the cost of leaving the operator sizes of the other half unchanged.

Following Eq. \eqref{eqn:norm_to_size}, the Pauli weight is given by (see the Supplemental material \cite{supp} section for a detailed derivation)
\begin{align}
\label{eqn:exact}
w(\hat{O})_{\text{ct}}&=\frac{1}{3^k}\sum_{N_{XY}\in\text{even}}\left(\begin{matrix}k\\N_{XY}\end{matrix}\right)2^{N_{XY}}\frac{1}{3^k}\nonumber\\
&+\frac{1}{3^k}\sum_{N_{XY}\in\text{odd}}\left(\begin{matrix}k\\N_{XY}\end{matrix}\right)2^{N_{XY}}\frac{1}{3^{N_{XY}}}\nonumber \\
&=\frac{1}{2}\Bigg[\frac{1}{3^k}+\frac{(-1)^{k}}{9^k}\Bigg]+\frac{1}{2}\Bigg[\left(\frac{5}{9}\right)^k-\frac{1}{9^k}\Bigg]. 
\end{align}
Note that the first and the second terms in $[...]$ of Eq. \eqref{eqn:exact} originate from the even and odd $N_{XY}$, respectively. In the large-$k$ limit, $w(\hat{O})_{\text{ct}}$ approaches zero, and the dominant term is $1/(2\times 1.8^k)$. Recalling that the shadow norm is related to $w(\hat{O})_{\text{ct}}$ by $\Vert\hat{O}\Vert^2_{\text{ct}} =w(\hat{O})_{\text{ct}}^{-1}$, we obtain the central result of this work $\Vert\hat{O}\Vert^2_{\text{ct}} \sim 2\times 1.8^k$. This demonstrates that the contractive unitary can outperform the random Clifford by significantly reducing the sample complexity for a subsystem with large-$k$. The key contribution to this improved scaling comes from the odd $N_{XY}$, which corresponds to the size distribution peaked at $2/3$. This reveals an important lesson: the ability to contract part of the operators to a smaller size improves the efficiency of classical shadow tomography.

\begin{figure*}
    \begin{center}  
    \includegraphics[width=0.7\linewidth]{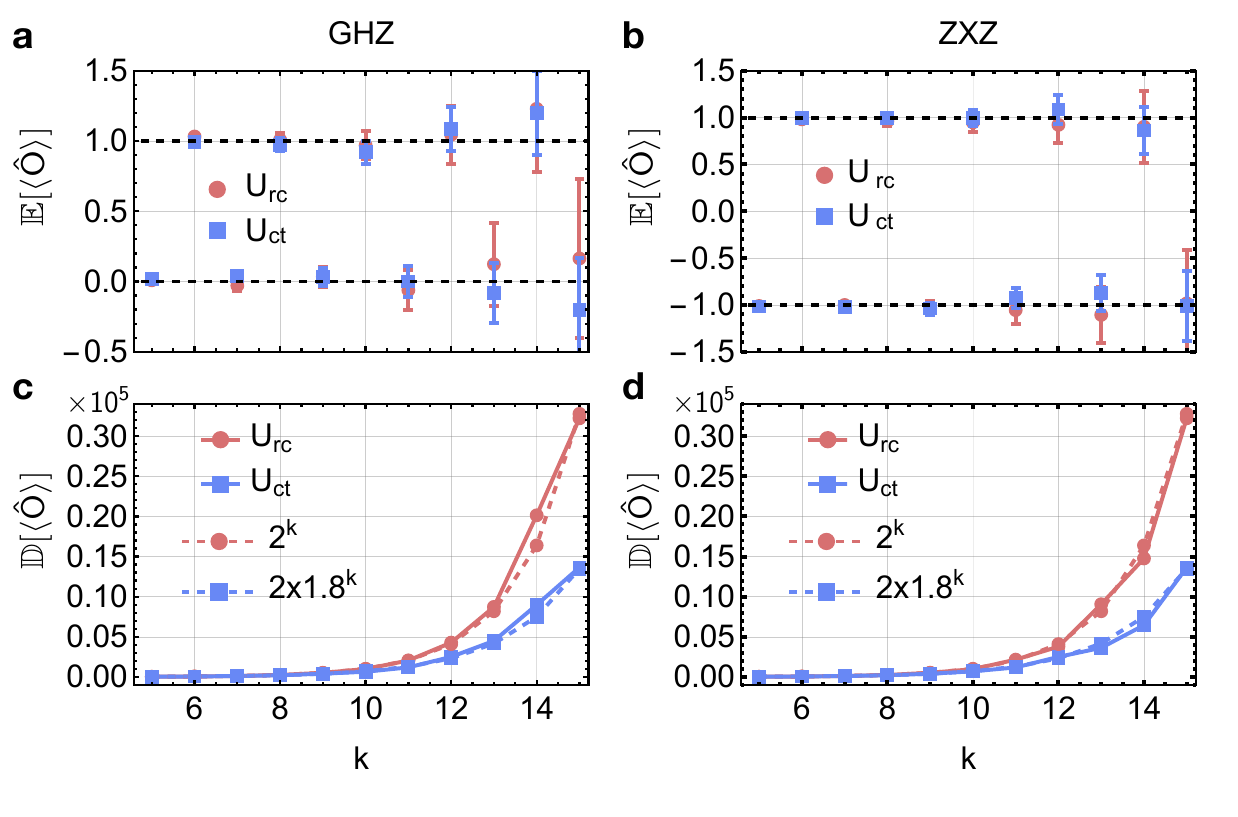}
    \caption{ \textbf{Numerical comparison between the random Clifford and contractive unitary protocols.} \textbf{a-b}, The expectation values of Pauli string operators predicted by classical snapshots. $\hat{Z}_1\cdots\hat{Z}_k$ is chosen for the GHZ state (a) and $\hat{Z}_1\hat{Y}_2\hat{X}_3\hat{X}_4\cdots\hat{X}_{k-2}\hat{Y}_{k-1}\hat{Z}_k$ is chosen for the ZXZ state (b), with different $k \in \{5,6,...,15\}$ within a system size of $N=20$. The error bars represent the standard deviation from $10^5$ classical snapshots. Black dashed lines indicate the exact values of the operator expectations. \textbf{c-d}, We compare the variance of the operator expectation predicted from different snapshots (solid lines) with the $2^k$ and $2 \times 1.8^k$ scaling laws for both protocols (dashed lines) derived from theoretical analysis. This demonstrates the advantage of our contractive unitary protocol at large $k$. } 
    \label{Fig:example}
    \end{center}
\end{figure*}

The discussion above focuses on operators with a size precisely equal to $k$. If there are $q$ identity operators $\hat{I}$ inserted into the Pauli string operator, then, following the same reason discussed above, these identity operators are converted to $\hat{Z}$ operators when $N_{XY}$ is odd. This contributes a coefficient of $1/3^q$ in front of the second term in Eq. \eqref{eqn:exact}. However, as long as $q\sim O(1)$ and does not scale with $k$, this coefficient only modifies the prefactor and does not change the $\sim 1.8^k$ scaling.

\textbf{Numerical Illustrations.} We further demonstrate our protocol with two numerical examples using $N$-qubit long-range entangled states: the Greenberger-Horne-Zeilinger (GHZ) state \cite{greenberger2007} and the one-dimensional cluster (ZXZ) \cite{NIELSEN2006147} state with periodic boundary conditions. We select the subsystem as $k$ successive qubits and make predictions for the string operator $\hat{O}=\hat{Z}_1\hat{Z}_2...\hat{Z}_{k-1}\hat{Z}_k$ for the GHZ state and $\hat{O}=\hat{Z}_1\hat{Y}_2\hat{X}_3\hat{X}_4...\hat{X}_{k-2}\hat{Y}_{k-1}\hat{Z}_k$ for the ZXZ state. Since both states admit efficient representations under the stabilizer formalism, the expectations of these operators can be derived analytically: $\langle\hat{O}\rangle=((-1)^k+1)/2$ for GHZ states and $\langle\hat{O}\rangle=(-1)^k$ for ZXZ states. These values serve as rigorous benchmarks.

In each sampling process, we first independently generate single-qubit rotations from 24 single-qubit Clifford gates. Then, we apply the composite unitary in Fig. \ref{Fig:illustration}\textbf{a} and sample the measurement outcome $\mathbf{z}^a$ in the computational basis. The prediction of the snapshot is computed by $O^a=\Vert\hat{O}\Vert_{\mathcal{E}_{U}}^2\text{Tr}(\hat{O} \sigma_U(\mathbf{z}^a))$ using the exact shadow norm in Eq. \eqref{eqn:norm_to_size}. After collecting $\mathcal{N}=10^5$ snapshots, the final prediction is made as a sample average $\mathbb{E}[\langle \hat{O} \rangle]\equiv \sum_{a=1}^{\mathcal{N}}O^a/\mathcal{N}$. The standard deviation of the expectation is estimated by $\sqrt{\mathbb{D}[ \langle \hat{O} \rangle]/\mathcal{N}}$. Here, $\mathbb{D}[ \langle \hat{O} \rangle]$ is the variance of samples, expected to match the shadow norm $\Vert\hat{O}\Vert_{\mathcal{E}_{U}}^2$. The results with system size $N=20$ are presented in Fig. \ref{Fig:example}, comparing the contractive unitary and random Clifford protocols. Fig. \ref{Fig:example}\textbf{a} and \textbf{b} show that both protocols provide an unbiased prediction for these string operators, while the contractive unitary protocol exhibits a smaller standard deviation. This is further supported by the plots of variances $\mathbb{D}[ \langle \hat{O} \rangle]$ in Fig. \ref{Fig:example}\textbf{c} and \textbf{d}, which directly confirm $2\times 1.8^k$ and $2^k$ scaling law shown by dashed lines.

\begin{figure}
    \begin{center}  
    \includegraphics[width=0.8\linewidth]{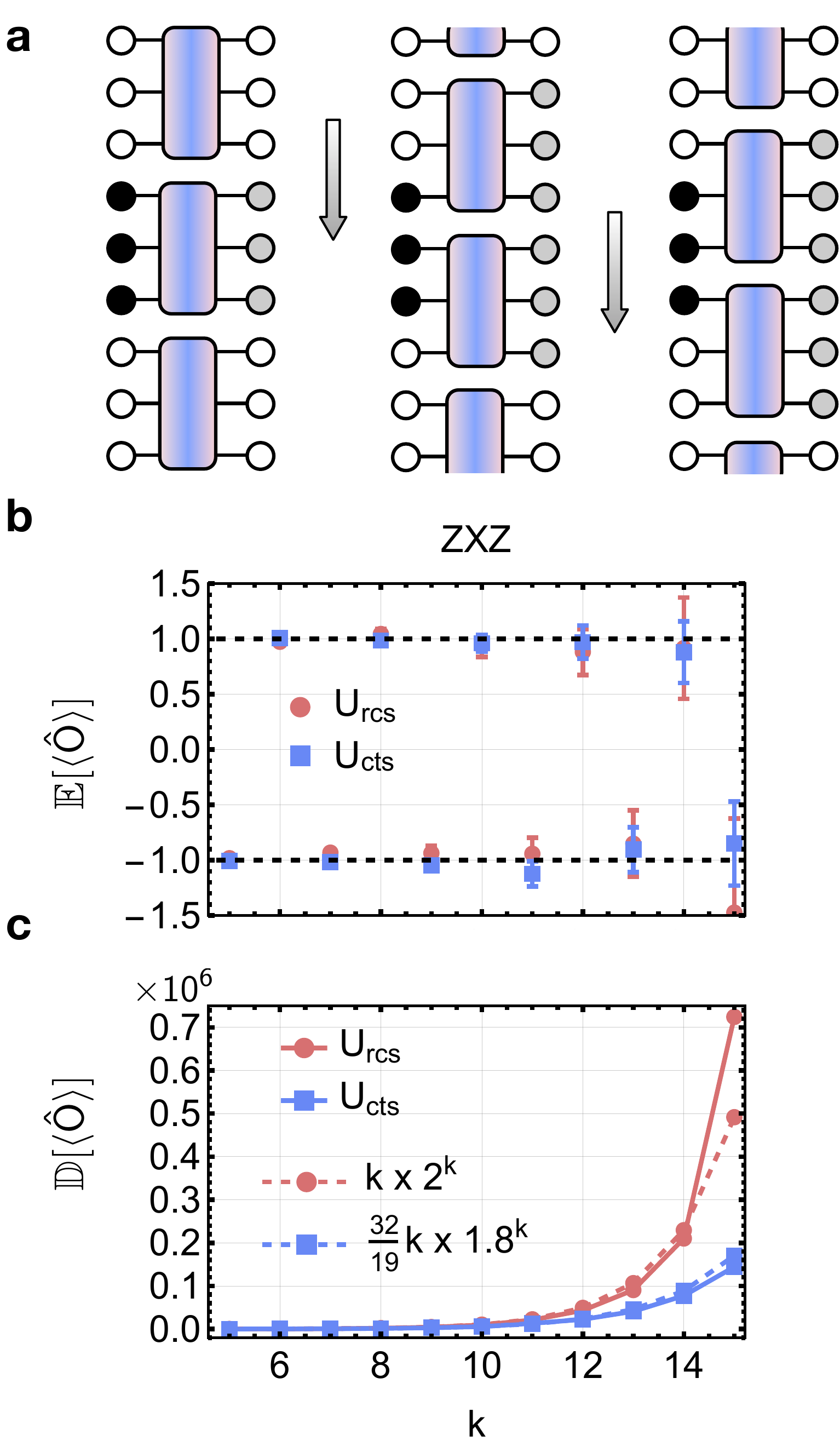}
    \caption{ \textbf{The sliding trick for situations in which the location of the Pauli string operator is unknown.} \textbf{a}, Each box represents an independent composite unitary applied to a subsystem with $k$ qubits, as shown in Fig. \ref{Fig:illustration}\textbf{a}. Two neighboring columns differ by sliding the unitary by one qubit. All these columns together form a complete sampling of unitaries. We randomly select each column with a probability of $1/k$. \textbf{b}, The expectation of the Pauli string operator $\hat{O}=\hat{Z}_{n_r+1}\hat{Y}_{n_r+2}\hat{X}_{n_r+3}\hat{X}_{n_r+4}\cdots\hat{X}_{n_r+k-2}\hat{Y}_{n_r+k-1}\hat{Z}_{n_r+k}$ for the ZXZ state. Here, $n_r\in[0,N)$ is a random integer. We vary $k$ in the range $k \in \{5,6,...,15\}$ with $N=3k$. The error bars represent the standard deviation calculated from $10^6$ snapshots. \textbf{c}, Solid lines denote the variance of the operator expectation predicted from classical snapshots, while dashed lines represent the theoretical scaling of $\sim (32/19)k\times 1.8^k$ and $k\times 2^k$ for these two protocols. } 
    \label{Fig:sliding}
    \end{center}
\end{figure}

\textbf{Extensition to Situations without Knowing Operator Locations.} The discussion above requires knowing the location where the operators act as a prior. If this regime is unknown, a naive extension requires applying the unitary on the entire system, yielding a shadow norm $\sim 2^N$ for the random Clifford protocol, with $N$ being the total number of qubits of the system. It results in an even larger shadow norm if the contractive unitary is applied to the entire system because it can convert the identity operator in the rest of the system back to $\hat{Z}$. There also exists another shallow circuits protocol that gives $\Vert\hat{O}\Vert^2_{\text{sc}}\gtrsim 2^k$ without prior knowledge of the location \cite{PhysRevLett.130.230403}. Here, we introduce a sliding trick, as shown in Fig. \ref{Fig:sliding}{\bf a}. With this trick, we can achieve the scaling of $\sim k\times 2^k$ and $\sim k\times 1.8^k$ for the random Clifford protocol and contractive unitary protocols, respectively. Although the random Clifford case still can hardly outperform the shallow circuit protocol, the contractive unitary protocol can for large enough $k$. 

To explain the sliding trick, we consider $N=n_0 k$ and arrange all qubits into a one-dimensional chain with the periodic boundary condition. We divide the whole system into $n_0$ subsystems, each containing $k$ qubits. We then apply unitaries sampled independently within each subsystem. Shadow tomography can predict the Pauli string operator efficiently if it lies within one of the subsystems, costing the same sample complexity as discussed above. However, it does not work well if the string operator crosses different subsystems. To overcome this difficulty, we slide all $n_0$ unitaries along one direction by one qubit, which generates another set of unitaries. Repeating this sliding can generate a total number of $k$ different sets of unitaries, as shown in Fig. \ref{Fig:sliding}{\bf a}. When performing the shadow tomography, we first randomly select a circuit structure with a probalility of $1/k$. By this construction, any given string operator can become compatible with the circuit structure with a probability of $1/k$. Hence, the sample complexity is bounded by $k\times 2^k$ and $2k\times 1.8^k$ for the random Clifford and the contractive unitary protocols, respectively. This upper bound assumes that once a string operator crosses two subsystems, the sample does not contribute at all. However, in practice, this is not strictly the case. By accounting for these contributions, more precise results for the contractive unitary protocol are $\sim (32/19)k\times 1.8^k$, while the random Clifford protocol maintains the same scaling. We present the exact results in the Supplemental material \cite{supp}.

We also present a numerical example in Fig. \ref{Fig:sliding} {\bf b} and {\bf c}. The example involves predicting the string operator $\hat{O}=\hat{Z}_{n_r+1}\hat{Y}_{n_r+2}\hat{X}_{n_r+3}\hat{X}_{n_r+4}\cdots\hat{X}_{n_r+k-2}\hat{Y}_{n_r+k-1}\hat{Z}_{n_r+k}$ for the ZXZ state. But now, $n_r$ is a random integer sampled from the range $[0, N)$, so the exact location of this string operator is unknown. We follow the sliding trick and use the exactly calculated shadow norm for this case to predict the expectation values. Fig.\ref{Fig:sliding}{\bf b} and {\bf c} show the numerical results for the expectation and the variance of this prediction, with a system size of $N=3k$. Equipped with the sliding trick, both the contractive unitary and random Clifford protocols can provide unbiased predictions for operator expectations with random locations. The dashed lines indicate that the variance for the contractive unitary and the random Clifford protocols is consistent with $(32/19) k\times1.8^k$ and $k\times 2^k$ scalings, respectively. 

\begin{table}[t]
  \centering
  \begin{tabular}{|c|c|c|c|}
    \hline
   \multirow{1.3}{*}{ Information of the  } & \multirow{1.3}{*}{ Contractive } & \multirow{1.3}{*}{ Random }  & \multirow{1.3}{*}{ Shallow }   \\
   \multirow{-1.3}{*}{ precise location of $\hat{O}$ }  & \multirow{-1.3}{*}{Unitary} & \multirow{-1.3}{*}{Clifford} & \multirow{-1.3}{*}{Circuit}  \\
    \hline
    Known& $1.8^k$ &  $2^k$ & \multirow{2}{*}{$> 2^k$} \\
    \hhline{---~}  
        Unknown     & $k\times 1.8^k$ &  $k\times 2^k$ &   \\
    \hline
  \end{tabular}
  \caption{A comparison of the sample complexity for the contractive unitary protocol, the random Clifford protocol, and the shallow circuits protocol for situations with or without the information of the precise location of the Pauli string operators $\hat{O}$.}
  \label{Table}
\end{table}

\textbf{Summary and Outlook.} We summarize our results in Table \ref{Table}. The key message is that, by utilizing the contractive unitary developed here, we can now reduce the sample complexity to $1.8^k$ or $k\times 1.8^k$ scaling for predicting Pauli string operators for situations with or without prior knowledge of the operator location. In contrast, previous results in both situations are always bounded by $2^k$ scaling, reached by random Clifford and shallow circuit protocols, respectively. Additionally, the contractive unitary also achieves low circuit complexity, as it only relies on deterministic and mutually commuting two-qubit gates. Hence, our proposal represents a major development in classical shadow tomography.

Before ending, we hope to highlight that the contractive unitary we proposed aligns perfectly with recent developments in the atom array quantum computation platform \cite{Bekenstein:2020aa,science.abg2530,doi:10.1126/science.abo6587,Bluvstein:2022aa,Evered:2023aa,Ma:2023aa,PhysRevX.13.041035,Bluvstein:2024aa,manetsch2024,PhysRevLett.133.013401,cao2024}. Firstly, the product of all $\hat{U}_{ij}$ connects every pair of qubits and therefore does not possess locality. This poses a challenge for other qubit platforms, although not impossible \cite{decross2024}, the reconfigurable nature of the atom array makes it feasible. Secondly, since all $\hat{U}_{ij}$ have identical forms, it requires at most $k-1$ steps of physical operations to implement this unitary, thanks to the ability to make parallel gate operations in atom array platforms. Finally, we note $\hat{U}_{\text{ct}}$ can be represented by elementary gates as $\hat{U}_{\text{ct}}=\prod_{i<j}\hat{U}^{\text{CZ}}_{ij}\prod_i\hat{S}^{k-1}_i$, where $\hat{U}^{\text{CZ}}_{ij}$ is a two-qubit CZ gate and $\hat{S}$ is a single-qubit phase gate. This single-qubit gate does not affect the contractive properties thus we can ignore it. Recent experiments on atom array platforms have achieved single-qubit rotation gate fidelity greater than $99.97\%$, and CZ gate fidelity $99.5\%$ \cite{Ma:2023aa,Evered:2023aa} and the global CZ gate has recently been realized \cite{cao2024}. Hence, the contractive unitary is readily implementable on the atom array quantum processor. 

This general idea of purposely designing deterministic quantum circuits that can efficiently contract or scramble operator size may find applications in quantum teleportation, quantum sensing, and quantum machine learning \cite{Biamonte:2017aa,RevModPhys.89.035002,Hu:2023aa}. Such designs typically require an all-to-all connection of qubits. With continuous advancements in the flexibility of controlling quantum devices, such as in the atom array platform, these theoretical approaches, as presented in this work, are increasingly demanding and practical.

\textit{Acknowledgments.}
We thank Yingfei Gu, Tian-Gang Zhou, and Huangjun Zhu for their helpful discussions.
This project is supported by HK GRF grants No. 17304820 (YW) and No. 17313122 (YW), CRF grants No. C6009-20G (YW) and No. C7012-21G (YW), a RGC Fellowship Award No. HKU RFS2223-7S03 (YW), the Science, Technology and Innovation Commission of Shenzhen, Municipality grants No. KQTD20210811090049034 (JY), Guangdong Basic and Applied Basic Research Foundation grant No. 2022B1515120021 (JY), the Innovation Program for Quantum Science and Technology 2021ZD0302005 (HZ), the XPLORER Prize (HZ), the NSFC under grant numbers U23A6004 (HZ), 12204352 (CW), and 12374477 (PZ), and a startup fund from UCSD (YZY).

\bibliography{refshadowZZ}

\end{document}


\title{Supplementary Material for ''Contractive Unitary and Classical Shadow Tomography''}

\author{Yadong Wu}
\affiliation{College of Physics, Sichuan University, Chengdu 610065, China}
\affiliation{Department of Physics and Hong Kong Institute of Quantum Science and
Technology, The University of Hong Kong, Pokfulam Road, Hong Kong SAR, China}

\author{Ce Wang}
\affiliation{School of Physics Science and Engineering, Tongji University, Shanghai, 200092, China}

\author{Juan Yao}
\affiliation{Shenzhen Institute for Quantum Science and Engineering, Southern University of Science and Technology, Shenzhen 518055, Guangdong, China}
\affiliation{International Quantum Academy, Shenzhen 518048, Guangdong, China}
\affiliation{Guangdong Provincial Key Laboratory of Quantum Science and Engineering, Southern University of Science and Technology, Shenzhen 518055, Guangdong, China}

\author{Hui Zhai}
\affiliation{Institute for Advanced Study, Tsinghua University, Beijing 100084, China}
\affiliation{Hefei National Laboratory, Hefei 230088, China}

\author{Yi-Zhuang You}
\affiliation{Department of Physics, University of California at San Diego, La Jolla, CA 92093, USA}

\author{Pengfei Zhang}
\email{pengfeizhang.physics@gmail.com}
\affiliation{Department of Physics \& State Key Laboratory of Surface Physics, Fudan University, Shanghai 200438, China}
\affiliation{Shanghai Qi Zhi Institute, AI Tower, Xuhui District, Shanghai 200232, China}

\date{\today}

\maketitle

\section{Shadow norm formula for Pauli operators}
 
In this section, we provide a review of the shadow norm of Pauli operators in locally-scrambled unitary ensemble for qudit systems, following the analysis in \cite{PhysRevLett.130.230403,PhysRevResearch.5.023027,Quantum.7.1026,npjQI.10.6,PhysRevLett.133.020602}. After the composite unitary operation $\hat{U}=(\prod_i \hat{u}_{2,i})~\hat{U}_g~(\prod_j \hat{u}_{1,j})$, the subsystem is measured in the computational basis. Each set of measurement outcome $|\mathbf{z}\rangle $ with $z_j\in\{0,1\}$ and $j\in\{1,2,...,k\}$ offers an classical snapshot $\hat{\sigma}_U(\mathbf{z})=\hat{U}^\dagger |\mathbf{z}\rangle \langle \mathbf{z}|\hat{U}$ of the quantum state. The average of classical snapshots defines a measurement channel, denoted as $\hat{\sigma} = \mathcal{M}[\hat{\rho}]_{\mathcal{E}_U}$, which reads 
\begin{equation}
\mathcal{M}[\hat{\rho}] _{\mathcal{E}_U}= \int_{\mathcal{E}_U}\textrm{d}\hat{U}\sum_{\mathbf{z}}\hat{U}^\dagger|\mathbf{z}\rangle\langle \mathbf{z}|\hat{U}~\langle \mathbf{z}|\hat{U}\hat{\rho}\hat{U}^\dagger|\mathbf{z}\rangle.
\end{equation}
The reconstruction of the original density matrix then requires the application of the inverse channel $\hat{\rho}=\mathcal{M}^{-1}[\hat{\sigma}]_{\mathcal{E}_U}$. With the locally-scrambled unitary ensemble, the measurement channel is diagonal in the Pauli string operator basis $\hat{O}_A$ applied to $k$ qudits: $\mathcal{M}[\hat{O}_A]_{\mathcal{E}_U} = w(\hat{O}_A)\hat{O}_A$, where $w(\hat{O}_A)$ is called the Pauli weight. Here, $A \in \{0, 1\}^{\otimes k}$ denotes the support of the Pauli string, namely $A_i = 1$ if there is a non-trivial single-qudit Pauli operator on site $i$. The size of the Pauli operator then corresponds to $m=\sum_i A_i$. With the knowledge of $w(\hat{O}_A)$, we can make predictions for Pauli observables using the fact that $\mathcal{M}$ (as well as its inverse) is self-adjoint:
\begin{equation}
\langle \hat{O}_A\rangle=\text{tr}(\hat{O}_A\mathcal{M}^{-1}[\hat{\sigma}]_{\mathcal{E}_U})=\text{tr}(\mathcal{M}^{-1}[\hat{O}_A]_{\mathcal{E}_U}\hat{\sigma})=w(\hat{O}_A)^{-1}\text{tr}(\hat{O}_A\hat{\sigma}).
\end{equation}

The remaining task is to compute the Pauli weight. By definition, we have:
 \begin{align}
 w(\hat{O}_A)=\frac{1}{D}\textrm{Tr}(\hat{O}_A^\dagger\mathcal{M}[\hat{O}_A]_{\mathcal{E}_U})&=\frac{1}{D}\int_{\mathcal{E}_U}\textrm{d}\hat{U}\sum_{\mathbf{z}}|\langle \mathbf{z}|\hat{U}\hat{O}_A\hat{U}^\dagger|\mathbf{z}\rangle|^2,=\int_{\mathcal{E}_U}\textrm{d}\hat{U}\textrm{d}\psi|\langle\psi|\hat{U}\hat{O}_A\hat{U}^\dagger|\psi\rangle|^2.\label{Seq.w1}
 \end{align}
Here, $D$ is the dimension of the Hilbert space. The basis $|\mathbf{z}\rangle$ can be replaced with a Haar-random product state, $|\psi\rangle = \otimes_{i=1}^k|\psi_i\rangle$, where each $|\psi_i\rangle$ is an independent Haar-random state. After unitary evolution, the Pauli operator $\hat{O}_A$ can be expanded in the Pauli basis $\hat{U}\hat{O}_A\hat{U}^\dagger = \sum_p\alpha_p\hat{O}_p$, where $\alpha_p = \textrm{Tr}(\hat{O}^\dagger_p\hat{U}\hat{O}_A\hat{U}^\dagger)/D$. 

For locally-scrambled ensembles, $\alpha_p$ obtains a random phase due to random single-qudit rotations. Therefore, the Pauli weight $w(\hat{O}_A)$ can be written as:
\begin{equation}
\begin{aligned}
w(\hat{O}_A)&=\int_{\mathcal{E}_U}\textrm{d}\hat{U}\textrm{d}\psi\sum_{p,p'}\alpha_p^*\alpha_{p'}\langle\psi|\hat{O}_p^\dagger|\psi\rangle\langle\psi|\hat{O}_{p'}|\psi\rangle,\\
&=\sum_{p,p'}\overline{\alpha_p^*\alpha_{p'}}\int\textrm{d}\psi\langle\psi|\hat{O}_p^\dagger|\psi\rangle\langle\psi|\hat{O}_{p'}|\psi\rangle,\\
&=\sum_p\overline{|\alpha_p|^2}\int\textrm{d}\psi | \langle\psi|\hat{O}_p|\psi\rangle|^2=\sum_p\overline{|\alpha_p|^2}\prod_{i=1}^k\int\textrm{d}\psi_i | \langle\psi_i|\hat{p}_i|\psi_i\rangle|^2.
\end{aligned}
\end{equation}
where $\hat{O}_p=\otimes_{i=1}^k\hat{p}_i$. For the single-qudit Haar-random state average:
\begin{align}
\int\textrm{d}\psi_i | \langle\psi_i|\hat{p}_i|\psi_i\rangle|^2=\left\{\begin{matrix}
1/(d+1),&\hat{p}_i \in \text{Pauli\ \ operator}\\
1,&\hat{p}_i \in \textrm{identity}
\end{matrix}\right.
\end{align}
By collecting all Pauli strings of the same Hamming weight in the sum 
\begin{align}
w(\hat{O}_A)=\sum_{m}^k(d+1)^{-m}\sum_{p:|p|=m}\overline{|\alpha_p|^2}\triangleq\sum_{m}^k(d+1)^{-m}\pi(m)\label{Seq.weight}.
\end{align}
where $\pi(m)$ is the Pauli operator size distribution after the evolution of the locally-scrambled unitary ensemble. We have $d=2$ for qubit systems.

The sample complexity for predicting observables depends on the variance of an operator's expectation after reconstruction, which is referred to as the shadow norm:
\begin{align}
\Vert\hat{O}_A\Vert^2_{\mathcal{E}_U}=\int_{\mathcal{E}_U}\textrm{d}\hat{U}\sum_{\mathbf{z}}\langle \mathbf{z}|\hat{U}\hat{\rho}\hat{U}^\dagger|\mathbf{z}\rangle|\textrm{Tr}(\hat{O}_A\mathcal{M}^{-1}[\hat{U}^\dagger|\mathbf{z}\rangle\langle \mathbf{z}|\hat{U}]_{\mathcal{E}_U})|^2.
\end{align}
For locally-scrambled unitary ensembles, the shadow norm of Pauli operators matches the inverse of the Pauli weight $\Vert\hat{O}_A\Vert^2 = w(\hat{O}_A)^{-1}$. This is because 
\begin{equation}
\begin{aligned}
\Vert\hat{O}_A\Vert^2_{\mathcal{E}_U}=&w(\hat{O}_A)^{-2}\int_{\mathcal{E}_U}\textrm{d}\hat{U}\sum_{\mathbf{z}}\langle \mathbf{z}|\hat{U}\hat{\rho}\hat{U}^\dagger|\mathbf{z}\rangle\langle \mathbf{z}|\hat{U}\hat{O}_A\hat{U}^\dagger|\mathbf{z}\rangle^2
\\=&D^{-1}w(\hat{O}_A)^{-2}\int_{\mathcal{E}_U}\textrm{d}\hat{U}\sum_{\mathbf{z}}\langle \mathbf{z}|\hat{U}\hat{O}_A\hat{U}^\dagger|\mathbf{z}\rangle^2\\=&D^{-1}w(\hat{O}_A)^{-2}\textrm{Tr}(\hat{O}_A^\dagger\mathcal{M}[\hat{O}_A]_{\mathcal{E}_U})=w(\hat{O}_A)^{-1}.
\end{aligned}
\end{equation}
Here, we observe that when expanding $\hat{\rho}=\hat{I}/D+\sum_{\hat{P}\neq I }c_{\hat{P}}\hat{P}$ over a complete basis of Pauli operators, only the identity part $\hat{I}/D$ contributes to the shadow norm. This is because for any $\hat{P}\neq \hat{I}$, we can always find another Pauli string $\hat{P}'$ such that $\{\hat{P},\hat{P}'\}=0$. The corresponding contribution vanishes, as replacing $\hat{U}\rightarrow \hat{U} \hat{P}'$ results in an overall minus sign. Consequently, if a locally-scrambled unitary ensemble $\mathcal{E}_U$ can contract the operator with size $k$ to a smaller size, then the probability $\pi(m)$ with $(m < k)$ increases, and the sample complexity would decrease.

 \vspace{5pt}

\section{Calculation of the Shadow Norm with the Contractive Unitary Evolution}

In this section, we will first present a proof that our result for the contractive unitary is one of the optimal two-qubit Clifford operators that approaches the minimal shadow norm of size-2 operators. Following that, we will provide details on calculating the shadow norm of size-$k$ Pauli operators after contractive unitary evolution, with knowledge of the subsystem's location where operators are applied. Next, we will modify the shadow norm of Pauli operators that contain identity operators in the subsystem. Finally, we will demonstrate the calculation of the shadow norm's coefficient in the sliding trick, without knowledge of the subsystem's location.

\subsection{Two-Qubit Contractive Unitary Operator}

Here, using commutation relation, we prove that a unitary operator can contract at most four size-2 Pauli operators to size-1 in the Clifford group. We arrange nine size-2 Pauli operators filled in this table $\mathcal{T}_2$ below:
\begin{table}[h]
\centering
\begin{tabular}{|c|c|c|} \hline 
XX & YX & ZX  \\ \hline
XY & YY & ZY  \\ \hline
XZ &	 YZ & ZZ  \\ \hline
\end{tabular}
\end{table}

In this table $\mathcal{T}_2$, three operators anti-commute with each other in every row and every column. When applying a unitary transformation to these operators, the commutation relations between them remain unchanged. There are six size-1 Pauli operators $\mathcal{P}_1=\{IX, IY, IZ, XI, YI, ZI\}$ for two-qubit system. Now we assume there are five size-2 operators that can be contracted by a unitary transformation $\hat{U}_c$ in the Clifford group. These five size-1 operators in $\mathcal{P}_1$ must contain three anti-commuting operators $IX, IY, IZ$ or $XI, YI, ZI$. Without loss of generality, we assume that the five size-1 operators are $\{IX, IY, IZ, AI, BI\}$ where $A, B$ denote individual Pauli  $X,Y,Z$ operators. $IX,IY,IZ$ anti-commute with each other so the their preimages must occupy one row(column) in the table $\mathcal{T}_2$. $AI$ and $BI$ commute with these three operators so their preimages can not share the same column(row) with any one of the preimages of $\{IX,IY,IZ\}$. However, this arrangement is impossible in a $3\times3$ table. Thus, we prove that at most four size-2 Pauli operators can be contracted.

\subsection{Shadow Norm for the size-$k$ Pauli Operators}

Here, we demonstrate the calculation details showing that the contractive unitary operator would yield a shadow norm with the scaling $\sim2 \times 1.8^k$. 

For the size-$k$ operator $\hat{O}_k$, under the single-qubit random Clifford unitary transformation $\prod_j\hat{u}_{1,j}$, there are $3^k$ different operators for $\hat{O} = \prod_j\hat{u}_{1,j}\hat{O}_k\hat{u}_{1,j}$, where each single-Pauli operator could be $X, Y, Z$ with equal probabilities. We denote the number of Pauli $X$ and $Y$ in $\hat{O}$ as $N_{XY}$. In the main text, we analyze how the contractive unitary $\hat{U}_{\rm ct}$ contracts the size-$k$ Pauli operator $\hat{O}$ to size-$N_{XY}$ if $N_{XY}$ is odd, while keeping the Pauli operator still size-$k$ with even $N_{XY}$. This leads to
\begin{equation}
m=\text{Size}\left(\hat{U}_{\rm ct} \hat{O}\hat{U}_{\rm ct}^\dagger\right)=
\begin{cases}
                   N_{XY} & \text{if}\ \   N_{XY} \in  \text{odd},  \\
                    k & \text{if}\ \   N_{XY} \in  \text{even}.
\end{cases}
\end{equation}
More precisely, the Pauli operator commutes with the contractive unitary with even $N_{XY}$, while anti-commutes with odd $N_{XY}$.

After the evolution, the Pauli weight depends on the size distribution from eq.(\ref{Seq.weight}). For an operator containing $N_{XY}$ $X$ and $Y$ operators, there are $\left(\begin{matrix}k\\N_{XY}\end{matrix}\right)2^{N_{XY}}$ configurations. Thus, the size distribution is:

\begin{equation}
\pi(m)_{\rm ct}=\delta_{m,k}~\frac{1}{3^k} \sum_{l=0} \left(\begin{matrix}k\\2l\end{matrix}\right)2^{2l}+
\begin{cases}
                  \frac{1}{3^k} \left(\begin{matrix}k\\m\end{matrix}\right)2^{m}& \text{if}\ \   m \in  \text{odd},  \\
                    0 & \text{if}\ \    \ \& \ \ m \in  \text{even},  
\end{cases}
\end{equation}

The Pauli weight can be obtained as:
\begin{align}
w(\hat{O}_k)_{\rm ct}=\sum_{m}\frac{\pi(m)_{\rm ct}}{3^m}=\frac{1}{3^k}\sum_{l=0} \left(\begin{matrix}k\\2l\end{matrix}\right)2^{2l}\frac{1}{3^k}+\frac{1}{3^k}\sum_{l=0} \left(\begin{matrix}k\\2l+1\end{matrix}\right)2^{2l+1}\frac{1}{3^{2l+1}}.\label{Seq:eo}
\end{align}
Inserting the equations
\begin{align}
\sum_{l=0} \left(\begin{matrix}k\\2l\end{matrix}\right)a^{2l}b^{k-2l}&=\frac{(a+b)^k+(-a+b)^k}{2},\nonumber\\
\sum_{l=0} \left(\begin{matrix}k\\2l+1\end{matrix}\right)a^{2l+1}b^{k-2l-1}&=\frac{(a+b)^k-(-a+b)^k}{2},\nonumber
\end{align}
into the eq.(\ref{Seq:eo}) we can obtain the Pauli weight:
\begin{align}
w(\hat{O}_k)_{\rm ct}&=\frac{1}{2}\frac{1}{9^k}(3^k+(-1)^k)+\frac{1}{2}\frac{1}{3^k}\left[\left(\frac{5}{3}\right)^k-\frac{1}{3^k}\right],\nonumber\\
&=\frac{1}{2}\left[\frac{1}{3^k}+\left(-\frac{1}{9}\right)^k\right]+\frac{1}{2}\left[\left(\frac{5}{9}\right)^k-\frac{1}{9^k}\right]\label{seq:ctscale}.
\end{align}
The shadow norm of a Pauli operator is the inverse of the Pauli weight, $\Vert\hat{O}_k\Vert^2 = w(\hat{O}_k)_{\rm ct}^{-1}$. For large $k$, the leading term of the shadow norm is $\sim2 \times 1.8^k$.

\subsection{Shadow Norm for the Operators containing Identity Operators }

Now, we discuss how the shadow norms are modified when the Pauli operator applied to the $k$-subsystem contains $q$ identity operators as defects. The size of the operator is $\tilde{k} = k - q$. After the single-qubit random unitary evolution, the size remains unchanged. The operator still contains $q$ identity operators. Once again, $N_{XY}$ denotes the number of Pauli $X$ and $Y$. If $N_{XY}$ is even, this operator commutes with the contractive unitary and cannot be contracted, maintaining a size of $\tilde{k}$. However, when $N_{XY}$ is odd, the $\tilde{k} - N_{XY}$ Pauli $Z$ operators will be converted to identity operators, while the $q$ identity operators will be converted to Pauli $Z$ operators. The Pauli weight will be modified as:

\begin{align}
w(\hat{O}_{\tilde{k},k})_{\rm ct}&=\frac{1}{3^{\tilde{k}}}\sum_{l=0} \left(\begin{matrix}\tilde{k}\\2l\end{matrix}\right)2^{2l}\frac{1}{3^{\tilde{k}}}+\frac{1}{3^{\tilde{k}}}\sum_{l=0} \left(\begin{matrix}\tilde{k}\\2l+1\end{matrix}\right)2^{2l+1}\frac{1}{3^{2l+1}}\frac{1}{3^{q}},\nonumber\\
&=\frac{1}{2}\left[\frac{1}{3^{\tilde{k}}}+\left(-\frac{1}{9}\right)^{\tilde{k}}\right]+\frac{1}{2}\left[\left(\frac{5}{9}\right)^{\tilde{k}}-\frac{1}{9^{\tilde{k}}}\right]\frac{1}{3^{q}},\label{Seq:ikd}\\
&=\frac{1}{2\times 3^{k-q}}+\frac{1}{2}\left(\frac{5}{9}\right)^{k}\left(\frac{3}{5}\right)^{q}+O\left(\frac{1}{9^k}\right).\nonumber
\end{align}
While there are $q\sim O(1)$ identity operators, for large $k$ the leading order of the shadow norm is $(5/3)^{q}\times2\times1.8^k$, which maintains the same scaling in $k$ as our main results.

If the number of identity operators scales linearly with $k$, that is, $q = \gamma k$, we can estimate the value of $\gamma$ such that the sample complexity of the contractive unitary is comparable to that of the Random Clifford unitary ensembles \cite{Nat.Phys.16.1050}, i.e., $(5/3)^{\gamma k} \times 1.8^k \approx 2^k$. In this case, $\gamma \approx 0.206$. The contractive unitary ensemble can predict the expectation with smaller sample complexity when non-trivial operators fill more than 80\% of the size of the $k$-Pauli strings compared to the random Clifford protocol. When the number of identity operators is much larger than the operator size $k$, the first term of Eq.\eqref{Seq:ikd} would be dominant and the shadow norm follows the scaling $\sim2 \times 3^k$.

\vspace{5pt}

\subsection{Shadow Norm of the Sliding Trick Protocol}

Here, we will present the calculation details regarding the shadow norm of the sliding trick protocol when we do not know the precise location where the Pauli operators are applied. For the simplest case, we discuss the entire system size $N = n_0k$, where $n_0$ is an integer. As per the sliding trick discussed in the main text, the $n_0$ unitary operators applied to the size-$k$ subsystem are independently sampled from the contractive unitary or the random Clifford ensembles. There are $k$ different circuit structures when these unitaries are applied by sliding one qubit. One of these circuits involves a unitary acting precisely on the subsystem where the Pauli operator is applied. However, every other $k-1$ circuit contains two unitaries acting on the subsystem where the Pauli operator is applied. The Pauli operator is divided into two parts with sizes $k_1$ and $k_2 = k - k_1$.

\begin{figure}[h]
        \centering
        \includegraphics[width=0.25\linewidth]{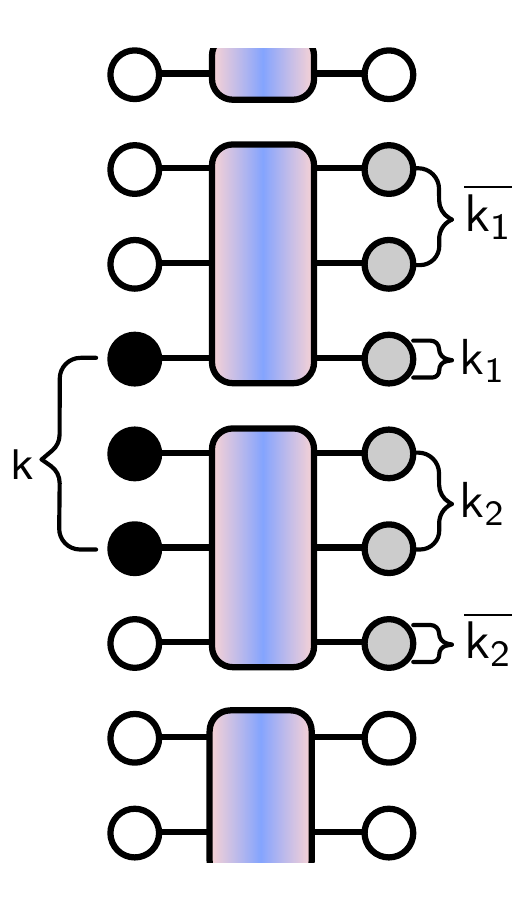}
        \caption{Sliding trick without knowledge of the precise location of size-$k$ operators. Two unitary operators are applied and dividing $k$ into $k_1$ and $k_2$ parts. Additional parts, $\overline{k_1}=k_2$ and $\overline{k_2}=k_1$, may be transformed into Pauli operators due to the global unitary.}
        \label{sfig:slid}
    \end{figure}
\vspace{5pt}

Assuming that the $k$-local operator is divided into two parts of size $k_1$ and $k_2=k-k_1$, denoted as $\hat{O}_{k_1}$ and $\hat{O}_{k_2}$ respectively, which are separated by two contractive unitary operators, as shown in FIG.\ref{sfig:slid}. If $\hat{O}_{k_1}$ ($\hat{O}_{k_2}$) contains an even number of Pauli $X$ and $Y$, it commutes with the contractive unitary and the operator size remains as $k_1$ ($k_2$). However, if there are an odd number of $X$ and $Y$ in $\hat{O}_{k_1}$ ($\hat{O}_{k_2}$), it can still be contracted, but the $\overline{k_1}$ ($\overline{k_2}$) part would be filled with Pauli $Z$ operators. Therefore, the Pauli weight of $\hat{O}_{k_1}$ after contraction can be written as this form according to eq.(\ref{Seq:ikd}).

\begin{align}
w(\hat{O}_{k_1,k})_{\rm ct}&=\frac{1}{2}\left[\frac{1}{3^{k_1}}+\left(-\frac{1}{9}\right)^{k_1}\right]+\frac{1}{2}\left[\left(\frac{5}{9}\right)^{k_1}-\frac{1}{9^{k_1}}\right]\frac{1}{3^{k-k_1}},\nonumber
\end{align}
Hence, the Pauli weight of the Pauli operator $\hat{O}_k$ can be expressed as:
\begin{align}
w(\hat{O}_k)_{\rm cts}&=\frac{1}{k}\sum_{k_1=1}^{k}w(\hat{O}_{k_1,k})_{\rm ct}w(\hat{O}_{k-k_1,k})_{\rm ct},\label{Seq:snorm}
\end{align}
where the coefficient $1/k$ indicates that each separation of $k_1=1,\cdots k$ can be sampled with equal probability $1/k$. The leading term of eq.(\ref{Seq:snorm}) is given by:
\begin{align}
\frac{1}{4k}\sum_{k_1=1}^{k}\Bigg[\left(\frac{5}{9}\right)^{k_1}\frac{1}{9^{k-k_1}}+\left(\frac{5}{9}\right)^{k_1}\left(-\frac{1}{27}\right)^{k-k_1}&+\left(\frac{5}{9}\right)^{k-k_1}\frac{1}{9^{k_1}}+\left(\frac{5}{9}\right)^{k-k_1}\left(-\frac{1}{27}\right)^{k_1}\Bigg].
\end{align}
Here the summation of the geometric sequence results in the leading term as:
\begin{align}
\frac{1}{4k}\left(\frac{5}{9}\right)^k\left(\frac{1}{4}-\frac{1}{16}+\frac{5}{4}+\frac{15}{16}\right)=\frac{19}{32k}\left(\frac{5}{9}\right)^k.
\end{align}
Therefore, by the definition of the shadow norm, $\Vert\hat{O}(k)\Vert^2_{\rm cts} =w(\hat{O}_k)_{\rm cts}^{-1}\sim(32/19)k\times 1.8^k$.

We can also apply the sliding trick to the random Clifford protocol, where each unit in FIG.\ref{sfig:slid} is independently sampled from a random Clifford group. After the unitary evolution, each site can be an identity or a Pauli $X,Y,Z$ operator with equal probabilities in both subsystems. In one subsystem, there are $4^k-1$ traceless Pauli operators. Therefore, the size distribution for each subsystem is given by $\pi(m)_{\rm rc}=\left(\begin{matrix}k\\m\end{matrix}\right)\frac{3^m}{4^k-1}$, where $m=1,\cdots,k$. So the Pauli weight of this random Clifford protocol with the sliding trick is given by:
\begin{align}
w(\hat{O}_k)_{\rm rcs}=&\frac{1}{k}\sum_{m=1}^k\left(\begin{matrix}k\\m\end{matrix}\right)\frac{3^m}{4^k-1}\frac{1}{3^m}+\frac{k-1}{k}\sum_{m_1=1}^k\sum_{m_2=1}^k\left(\begin{matrix}k\\m_1\end{matrix}\right)\frac{3^{m_1}}{4^k-1}\frac{1}{3^{m_1}}\left(\begin{matrix}k\\m_2\end{matrix}\right)\frac{3^{m_2}}{4^k-1}\frac{1}{3^{m_2}},\label{seq:slideRC}\\
&=\frac{1}{k}\frac{1}{2^k+1}+\frac{k-1}{k}\frac{1}{(2^k+1)^2}.\nonumber
\end{align}
The first term represents the size-$k$ operator that is not separated by two Clifford operators, with probability $1/k$. The second term represents the other $k-1$ different separations, which contribute less to the Pauli weight. Thus, the shadow norm scales with $k\times(2^k+1)$

We have also discovered that the scaling $k \times 1.8^k$ remains consistent when the system size cannot be perfectly divided into $k$ subsystems. Let us assume that the system size is $N = n_0k + q$, where $q = 1, \cdots, k-1$. We can use $n_0$ unitary operators, each applied to a size-$k$ subsystem, and one unitary applied to a size-$q$ subsystem. These $n_0+1$ unitary operators are independently sampled and slid $N$ times with periodic boundary conditions to construct the circuit ensemble using the sliding trick. Each circuit structure is sampled with a probability of $1/N$. The size-$k$ operator can be contracted by one unitary operator, which is precisely applied to the same subsystem. There are $n_0$ circuits containing this unitary operator. Therefore, the dominant term of the Pauli weight under this ensemble is $\frac{n_0}{N}\left(\frac{5}{9}\right)^k$, and the shadow norm maintains the scaling $k \times 1.8^k$ unchanged.


\bibliographystyle{unsrt}  
